\documentclass[twocolumn,showpacs,preprintnumbers,amsmath,amssymb,prb]{revtex4}

\usepackage{graphicx}
\usepackage{dcolumn}
\usepackage{bm}
\usepackage{times}
\usepackage{mathptm}
\begin{document}


\title{Drag force on an oscillating object in quantum turbulence}

\author{Shoji Fujiyama}
\author{Makoto Tsubota}
\affiliation{Graduate School of Science, Osaka City University, Osaka 558-8585, Japan}
\date{\today}

\begin{abstract}
This paper reports results of the computation of the drag force exerted on an oscillating object in quantum turbulence in superfluid $^4$He.
 The drag force is calculated on the basis of numerical simulations of quantum turbulent flow about the object.
The drag force is proportional to the square of the magnitude of the oscillation velocity, which is similar to that in classical turbulence at high Reynolds number.
The drag coefficient is also calculated, and its value is found to be of the same order as that observed in previous experiments.
The correspondence between quantum and classical turbulences is further clarified by examining the turbulence created by oscillating objects.

\end{abstract}

\maketitle

\section{introduction}\label{introduction}

In recent years, quantum turbulence has become an established research area in low-temperature physics.\cite{Tsubota,PLTP}
The similarities and differences of quantum turbulence in relation to classical turbulence comprise one area of interest of quantum turbulence.
For example, although both classical and quantum turbulences consist of eddies or vortices, these two kinds of turbulences are quite different in terms of their hydrodynamic properties.

First, classical fluid can have arbitrary vorticity, while in contrast the vorticity in superfluid $^4$He is quantized;
circulation around a quantized vortex is $\kappa = h/m$, where $h$ is Planck's constant and $m$ is mass of $^4$He atom, with the core size of a quantized vortex given by the healing length of superfluid $^4$He, which is about the atomic size of $^4$He.
Moreover, quantized vortices cannot be generated under usual experimental conditions, unlike eddies in classical fluids.
Instead, remnant vortices are essential for the generation of quantum turbulence.
It is believed that remnant vortices are usually pinned to the roughness of the wall.\cite{Donnelly}
Indeed, Hashimoto \textit{et al.}\cite{Hashimoto} experimentally showed that quantum turbulence cannot be generated without remnant vortices.

Second, classical fluid is viscous, while superfluid $^4$He consists of viscous normal fluid and an inviscid superfluid according to a two-fluid model.
The two fluids are coupled in turbulence through mutual friction, which is the interaction between vortices and normal fluid.
The ratio of the two fluids is temperature dependent, and the normal fluid dominates the fluid if the temperature is close to the transition temperature, while
in the limit of zero temperature, the superfluid component dominates the fluid, and pure quantum turbulence is achieved. The attention of the present study is confined to this latter limiting case.

In spite of the above distinct properties, quantum and classical turbulences also exhibit similarities.
Kolmogorov's law, which is a statistical law found in a classical turbulence, has also been observed in quantum turbulence both experimentally and numerically.\cite{Nore,Maurer,Stalp,Araki1,Kobayashi}
This law can thus be taken to be universal over quantum and classical turbulences.

Under this new motivation, quantum turbulence has recently been studied, and accordingly the progress of experimental techniques has given rise to a variety of methods to create and observe quantum turbulence.
To both create and detect quantum turbulence, objects oscillating in a superfluid, such as spheres, grids, thin wires, and tuning forks are often employed.\cite{Hashimoto,Yano,Goto,Schoepe,McClintock,Fisher,Bradley,Blazkova}
The most commonly measured physical quantity is the velocity of the oscillating objects.
At temperatures lower than 1 K, the velocity exhibits the clear transition from laminar to turbulent state when the driving force is increased.\cite{Vinen}
For magnitudes of  velocities less than about 50 mm/s, the drag force is proportional to the oscillation velocity due to collisions between the oscillating object and some remaining excitations.
At higher  magnitudes of velocity, the drag force is proportional to the square of the oscillation velocity magnitude because turbulence is generated around the object.

These results also describe a similarity between quantum and classical turbulence;
in classical fluid, the drag force on an object in a uniform flow at high Reynolds number is described by
\begin{equation}
	F_D = \frac{1}{2} C_D \rho A U^2, \label{eqn:drag_force}
\end{equation}
where $C_D$ is the drag coefficient, $\rho$ is the density of the fluid, $A$ is the projection area of the object normal to the flow, and $U$ is the flow velocity.\cite{Vinen}
This relation could be applicable to quantum turbulence created by an oscillating object.
At low Reynolds number, Stokes's drag force acts on the object, which is proportional to the magnitude of the flow velocity, with the result that the drag coefficient $C_D$ becomes inversely proportional to the magnitude of the flow velocity.
Fitting experimental results to a relationship of the form of Eq.\ (\ref{eqn:drag_force}), we derive a constant drag coefficient, which is of order unity, as for classical turbulence.

In this paper, we shall further add to knowledge of the similarities of quantum and classical turbulences by presenting numerical simulations of the drag force and coefficient on the oscillating object.
However, in calculating the drag force numerically, the pressure on the object needs to be calculated over the entire surface of the object.
The pressure at each point on the object can be calculated from the velocity field by Euler's equation, and the velocity field is generated by the vortices.
It is possible to calculate the pressure if the number of the vortices is relatively small,\cite{Kivotides} but it is difficult with current computational resources to integrate the pressure in quantum turbulence in cases where a large number of vortices exist.
As a consequence, we shall introduce a method to evaluate the drag force.
We assume that the oscillating object does work on the fluid when the vortices are grown by the object and that a drag force is caused as a result of the work.
It is shown later that the drag force can be estimated from the energy given to the fluid by the object in an equilibrium state of quantum turbulence.
The increase in the kinetic energy of the fluid can be evaluated from the energy of vortices grown by the object.
Finally, from the energy given per unit time divided by the velocity of the object we derive the drag force.

In this paper, the drag force in quantum turbulence is targeted for the following reason:
the drag force becomes definite according to the method of this paper only if the vortices are grown by the oscillating object.
In the laminar regime, it is expected that the drag force is caused by collisions between the object and excitations and that the growth of vortices is not essential.
In the turbulent regime, on the other hand, the drag force is caused by the growth of vortices.
For this reason, we focus on the dynamics of vortices around an oscillating object and solve the equations of motion for the vortices numerically.

In the simulation, we consider an  object oscillating with a constant magnitude of velocity $v$, and estimate the drag force due to the vortices, that is the drag force $F_\mathrm{drag}$ is described in the form $F_\mathrm{drag} = f(v)$.
Note that this is in contrast to experimental studies, which measure the oscillation velocity of the object for a certain driving force, obtaining the velocity $v = f^{-1}(F_\mathrm{drag})$.
The functional relation $F_\mathrm{drag} = f(v)$ used in the present simulations is thus inverse to the relation used in the experiments, but it is worth comparing the results for the following reason.
As we shall see in Sec.\ \ref{turbulence}, when the object oscillates, remnant vortices are stretched and the vortex tangle grows, which results in the drag force on the object.
Finally, the vortex tangle reaches an equilibrium state of quantum turbulence.
The drag force is then uniquely determined by the oscillation velocity in the simulation.
In an experimental equilibrium state\cite{Yano,Schoepe,McClintock,Fisher,Bradley,Blazkova} on the other hand, the drag force responsible for the dissipation by vortices must be equal to the driving force that injects the energy into the system.
In addition, the oscillation velocity is measured under a constant driving force, namely, drag force in an equilibrium state of turbulence in the experiments.
Hence there is one-to-one correspondence between the oscillation velocity and the drag force, which allows us to compare the drag force of the simulation with that of the experiments.

The coupled dynamics of vortices and the oscillating object is not considered in this work, which is validated by verifying that the drag force depends only on the oscillation velocity.
The oscillation of the object causes the growth of vortices, and the rate of the growth depends on the oscillation velocity because the vortices attached to the object are stretched by the object (details in Sec.\ \ref{turbulence}).
 The vortex dynamics are influenced only by the magnitude of the oscillation velocity irrespective of whether the dynamics of  the object are considered or not.
This is because since the drag force is evaluated from the rate of the vortex growth, the drag force does not depend on the details of the motion of the object but the oscillation velocity.

A previous  numerical simulation of an oscillating sphere was performed by H\"anninen \textit{et al.}\cite{Hanninen1}
They used the same size of the sphere and frequency of the oscillation as the experiment of Shoepe \textit{et al.}\cite{Schoepe}
The simulation seems to show a growth of vortices toward turbulence, but an equilibrium state of quantum turbulence is not achieved, which differs from the present results.
This is because in H\"anninen \textit{et al.}'s\cite{Hanninen1} study the vortices extending between the wall and the sphere remained attached and continued to generate vortex rings, while in our simulations, mature vortices are soon detached from the object and the remaining vortices are able to grow successively (details in Sec.\ \ref{turbulence}).
It is important to note that the drag force cannot be calculated from the simulations of H\"anninen \textit{et al.}\cite{Hanninen1} in the manner of the present study, since our method assumes an equilibrium state.

The contents of this paper are as follows.
In Sec.\ \ref{formulation}, we shall clarify the configuration and formulation of  the model and introduce the equations of motion required in the dynamics of quantum turbulence.
In Sec.\ \ref{turbulence}, we shall describe the process whereby quantum turbulence is generated by an oscillating object.
In Sec.\ \ref{drag force}, we shall introduce the present proposed method of evaluating the drag force on the oscillating object and show the dependency of the drag force on the velocity of the object.
Section \ref{conclusion} is devoted to the conclusion and the discussion.

\section{model and formulation}\label{formulation}

Superfluid $^4$He at 0 K can be treated mathematically as an ideal incompressible fluid of vanishing viscosity.
In a superfluid, any circulation is quantized in units of $\kappa$.
The vorticity is infinitely distributed on thin vortex core, whose size is the order of 1 \AA, so that the circulation around the vortex filament is restricted to $\kappa$.\cite{Donnelly}
With these features, it is valid to adopt the vortex filament approximation, which regards a vortex as a line.
Numerical study of quantum turbulence can be accomplished by solving the equations of motion for the vortex filaments that form the turbulence.\cite{Schwarz85,Schwarz88}
According to Helmholtz's theorem, a vortex filament moves with superfluid velocity on the vortex.
In an infinite system of superfluid, the superfluid velocity is equal to the velocity generated by vortex filaments, which takes the form of the Biot-Savart integration,
\begin{equation}
	\bm{v_\omega}(\bm{r})=\frac{\kappa}{4\pi} \int _\mathcal{L}\frac{(\bm{s_1}-\bm{r}) \times d \bm{s_1}}{\left | \bm{s_1}-\bm{r} \right |^3},\label{eqn:Biot-Savart}
\end{equation}
where $\bm{s_1}$ is a position vector on a vortex and the integration is performed over all the vortex lines.
In attempting to determine a velocity field at a certain point $\bm{r} = \bm{s}$ on a vortex, Eq.\ (\ref{eqn:Biot-Savart}) diverges as $\bm{s_1} \rightarrow \bm{s}$.
The divergence is evaded by introducing cutoff parameter $a_0$ in Eq.\ (\ref{eqn:Biot-Savart}).

In the presence of a boundary, additional superflow $\bm{v_b}$ appears so as to satisfy the boundary condition for an inviscid fluid,
\begin{eqnarray}\label{boundary}
	&\bm{v_s} \cdot \bm{n} = 0,& \nonumber \\
	&\bm{v_s} = \bm{v_\omega}+\bm{v_b},&
\end{eqnarray}
where $\bm{n}$ is a vector normal to the boundary.
While many possible geometries may be considered for the vibrating objects, in the present study we restrict attention to the sphere for simplicity.

For a velocity field generated by an infinitesimal element $d\bm{s}$ of a closed-loop vortex, Eq.\ (\ref{boundary}) is satisfied by considering an image vortex inside the sphere.\cite{Saffman,Laplace_solve}
\begin{figure}[tb]
	\includegraphics[width=0.9\linewidth]{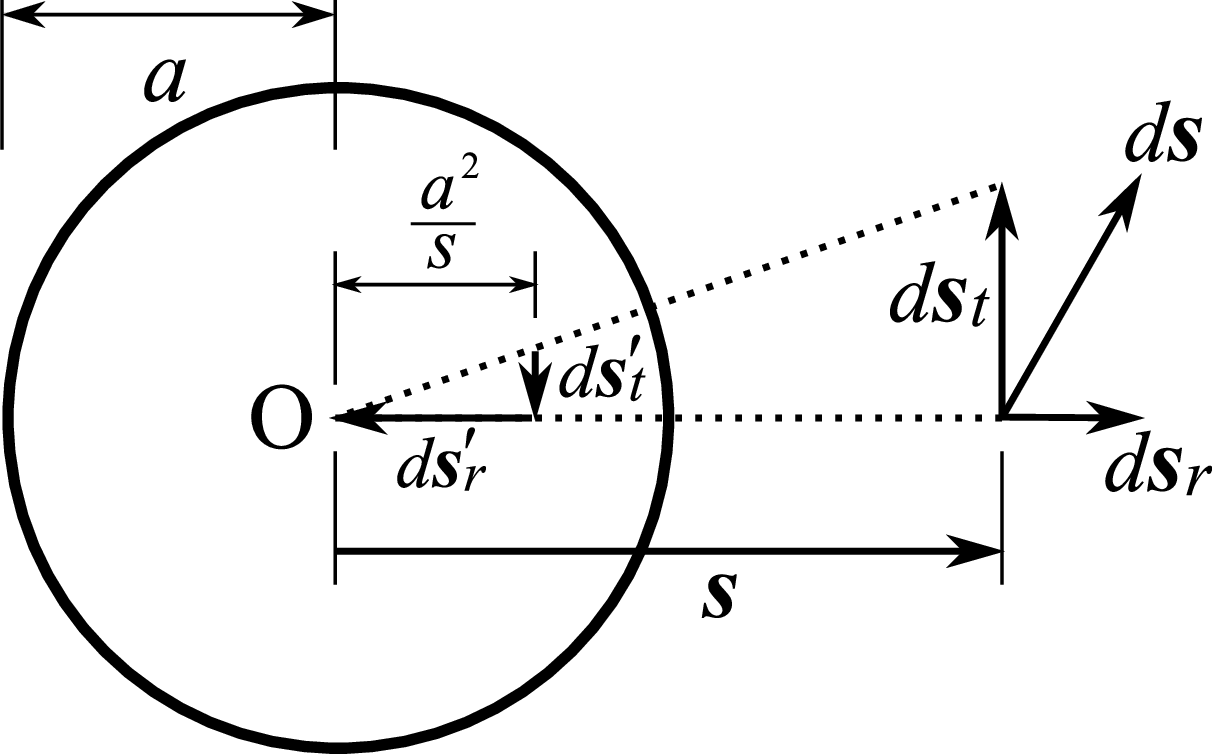}
	\caption{Image vortices in a sphere for a vortex element.}\label{fig:img_vor}
\end{figure}
The image vortex consists of two parts: a radial image $d\bm{{s'}_r}$ and a tangential image $d\bm{{s'}_t}$.
In Fig.\ \ref{fig:img_vor}, the origin is set at the center of the sphere, $a$ is the radius of the sphere, $\bm{s}$ is a radius vector of the vortex element, $d\bm{s_r}$ is a radial component of $d\bm{s}$, and $d\bm{s_t}$ is a tangential component defined as $d\bm{s_t} = d\bm{s} - d\bm{s_r}$.
The tangential image is located at $a^2/s$ from the origin and anti-parallel to $d\bm{s_t}$.
Its length is $(a/s)^2 ds_t$, and it has circulation $(s/a)\kappa$.
The radial image directs from $d\bm{s'_t}$ to the origin, its length is $a^2/s$ and it has circulation $(ds_r a)\kappa$.

When the sphere moves with the velocity $\bm{u_p}$, the boundary condition [Eq.\ (\ref{boundary})] is satisfied in a reference frame that moves with the sphere:
\begin{equation}
	(\bm{v_s}-\bm{u_p}) \cdot \bm{n} = 0. \label{eqn:boundary_with_sphere}
\end{equation}
The extra term in the parenthesis is canceled out by adding the following term to $\bm{v_b}$:
\begin{eqnarray}
	\bm{v_u}(\bm{r}) &=& \nabla \Phi _u (\bm{r}), \label{eqn:uniform_flow} \nonumber \\
	\Phi _u (\bm{r}) &=& -\frac{1}{2} \left(\frac{a}{r}\right)^3 \bm{u_p}\cdot\bm{r},
\end{eqnarray}
where $\bm{v_u}$ is a flow induced by the motion of the sphere and determined by the instantaneous velocity of the sphere.\cite{Schwarz74}
Thus $\bm{v_b}$ is obtained by adding the contribution from the image vortices and the velocity field given by Eq. (\ref{eqn:uniform_flow}).

The dynamics of the vortices are completely determined by the above formulation.
When two vortices cross or when a vortex becomes close to a spherical boundary, it is theoretically known that reconnection, which is topological change, occurs.\cite{Koplik}
In the framework of the vortex filament model, the details of the vortex core are neglected, and so these formulations do not describe the dynamics of the reconnection.
In the present numerical simulation, an exceptional routine is included in our numerical code to allow reconnection to occur when a vortex approaches another vortex or a spherical boundary within the numerical resolution.\cite{Araki2}

It is thought that quantum turbulence originates from remnant vortices rather than from the intrinsic nucleation of vortices.\cite{Hashimoto,Goto,Donnelly}
Thus we inject vortex rings toward the oscillating object to induce quantum turbulence, considering the experiment of the Osaka group.\cite{Goto}
 This approach using the injection of vortices requires the consideration of two parameters: the ring size and the time interval in which the vortices are injected.
The ring size can be estimated experimentally by the time of flight for the vortex ring, and we utilize the value of the ring size 1 $\mu$m.\cite{Goto}
We assume that the time interval is smaller than the period of the oscillation, say, 0.05 ms.\cite{Goto}
In order to replicate the conditions of Ref.\ 15, in the present study the same  parameter values as this previous study are used, such as the radius of sphere and the oscillation frequency of the sphere.
The diameter of the sphere is 3 $\mu$m, the frequency of the oscillation is 1590 Hz, while the oscillation velocity is chosen in the range of 30--90 mm/s.

\section{development of turbulence}\label{turbulence}
\begin{figure}[tb]
	\begin{tabular}{cc}
		\includegraphics[width=0.48\linewidth]{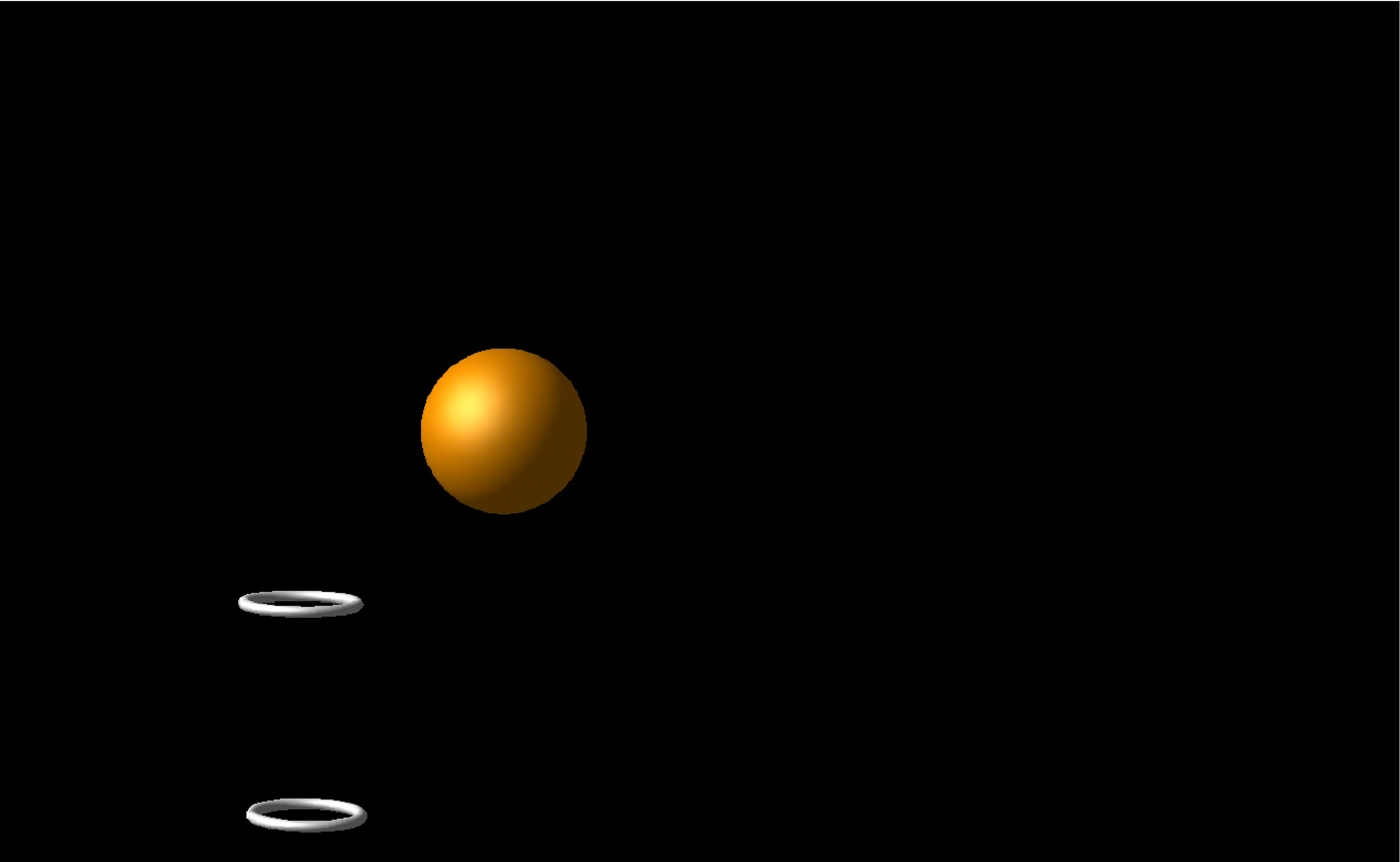} &
		\includegraphics[width=0.48\linewidth]{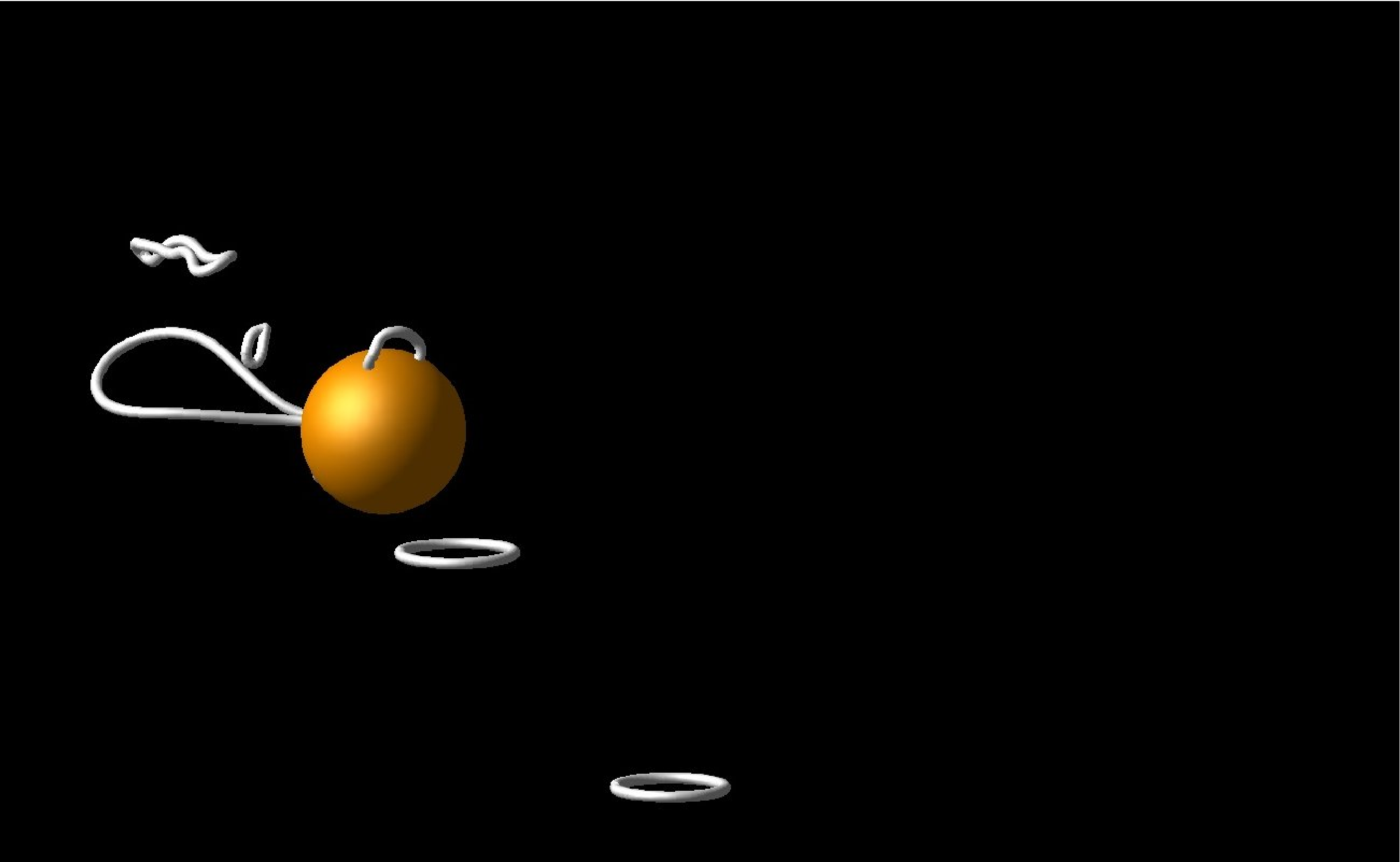} \\
		(a) $t$ = 0.19 ms & (b) $t$ = 0.40 ms \\
		\includegraphics[width=0.48\linewidth]{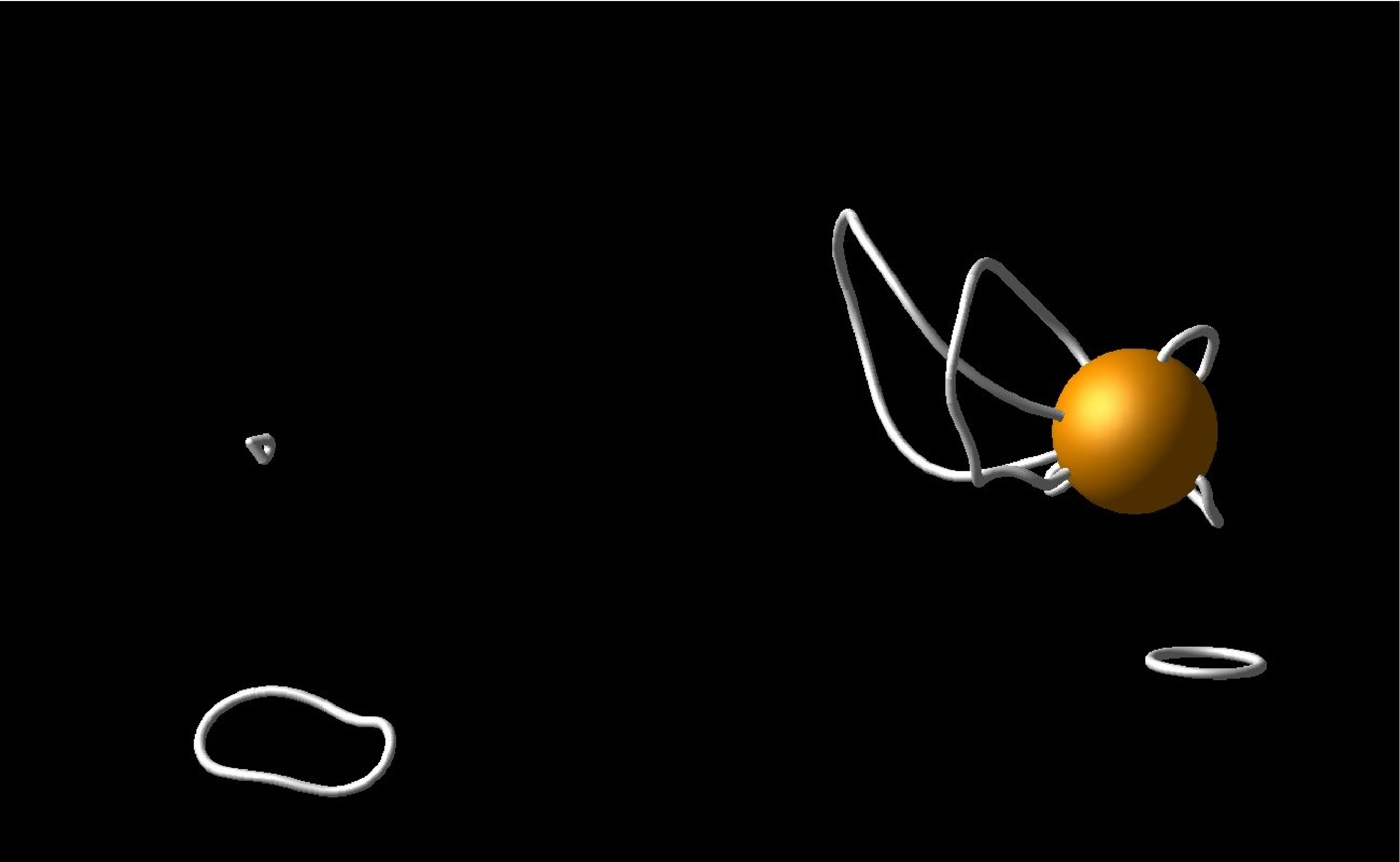} &
		\includegraphics[width=0.48\linewidth]{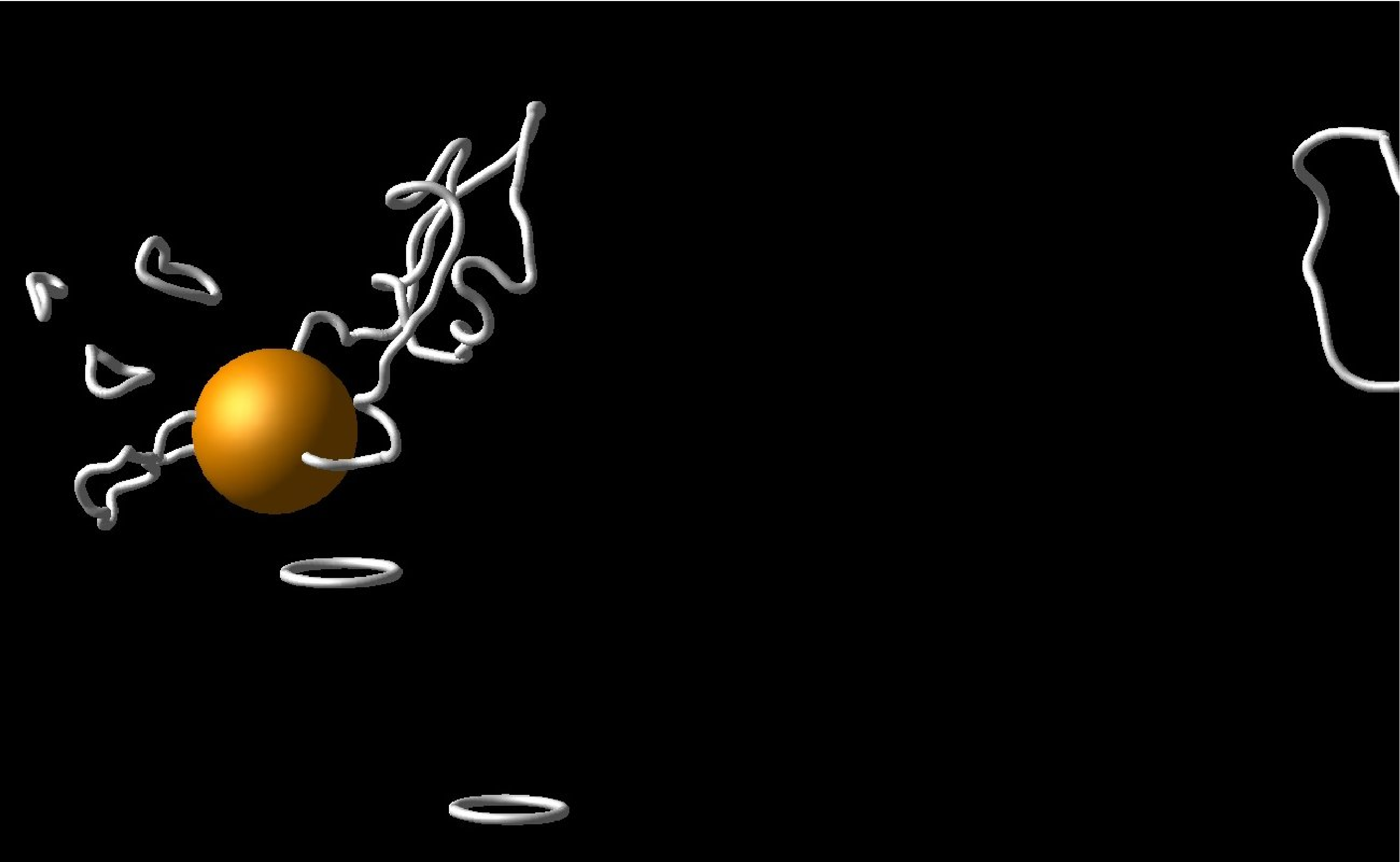} \\
		(c) $t$ = 0.58 ms & (d) $t$ = 1.00 ms
	\end{tabular}
	\caption{(Color online) The time evolution of turbulence generation for the case of a sphere oscillating with a velocity magnitude of  90 mm/s.
		See the text for further details.}\label{fig:turbulence}
\end{figure}
The time evolution of the turbulence generation is shown in Fig.\ \ref{fig:turbulence} and Ref.\ 27.
The sphere oscillates horizontally, and vortex rings are injected from the bottom of the medium [Fig.\ \ref{fig:turbulence}(a)].
When the vortex rings collide with the sphere, reconnection occurs and the vortices are attached to the sphere.
It can be seen that the attached vortices are stretched as the sphere moves [Figs.\ \ref{fig:turbulence}(b) and \ref{fig:turbulence}(c)].
Due to the successive injection of vortex rings, the process is repeated and the stretched vortices form a tangle around the sphere [Fig.\ \ref{fig:turbulence}(d)].
The vortices grow in size and are then detached from the sphere as follows.
The flow caused by the motion of the sphere [Eq.\ (\ref{eqn:uniform_flow})] drives the end points of the attached vortices to the stagnation point of the sphere.
A pair of the end points of the vortex approaches each other as they converge to the stagnation point.
Finally, reconnection of the pair of the end points then occurs and the vortex is detached from the sphere [Fig.\ \ref{fig:turbulence}(c)].
In spite of the detachment of the vortices, the oscillating sphere still sustains the vortex tangle.
\begin{figure}[tb]
	\includegraphics[width=\linewidth]{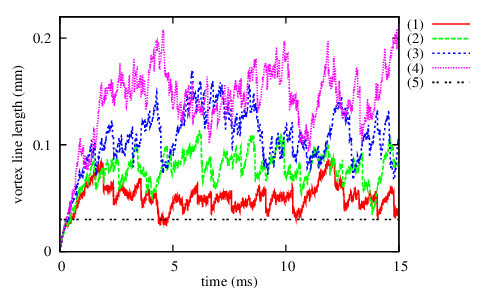}
	\caption{(Color online) Vortex line length at different velocities of the oscillation.
		(1) 30 mm/s, (2) 50 mm/s, (3) 70 mm/s, (4) 90 mm/s, and
		(5) vortex line length in the absence of the sphere.}\label{fig:length}
\end{figure}

Figure \ref{fig:length} shows the vortex line length at different oscillation velocities as a function of elapsed time.
Only the vortex line length inside the computational box (40 $\mu$m)$^3$ is calculated.
In Fig.\ \ref{fig:length}, the line length increases in the first 0.5 ms as the vortex rings are injected.
After the vortices are attached, the vortices are stretched and correspondingly the line length continues to increase.
However, when detached vortices escape from the computational box, the line length suddenly drops.
The loss of the vortices balances out the injection and the growth of the vortices, so that the line length saturates.
Although a slight increase in the line length can be seen for a velocity magnitude of 30 mm/s, the line length almost lies on line (5) in Fig.\ \ref{fig:length}, which means that the vortices are not stretched by the sphere.
It can be seen that the saturated value of the line length increases as the oscillation velocity magnitude increases.
 For velocity magnitudes above 50 mm/s the saturated line length value is larger than the injection of vortices, which suggests that vortex tangles are forming around the sphere.

\section{drag force}\label{drag force}
In this section, we propose a method to numerically evaluate the drag force exerted on a sphere oscillating in quantum turbulence, and calculate it as a function of the magnitude of the oscillation velocity.

A sphere with acceleration $d\bm{u_p}/dt$ experiences a force exerted by the fluid given by\cite{Landau}
\begin{equation}
	F_d = -\frac{2}{3}\pi a^3\rho_s \frac{d\bm{u_p}}{dt}. \label{eqn:Fd}
\end{equation}
Furthermore, in quantum turbulence, an extra drag force acts on the sphere for the following reason.
In the equilibrium state of quantum turbulence, the injected vortex rings are constantly grown by the sphere.
The sphere pushes the fluid and does work on the fluid because the kinetic energy of the fluid increases due to the stretching of the vortices, and a consequent reaction force acts on the sphere.
The force $F_d$ described above is the order of 0.01 pN with the parameters used in this paper, being found negligible compared with the additional quantum drag force 1 pN, obtained later of the order of 1 pN.

\begin{figure}[tb]
	\includegraphics[width=0.9\linewidth]{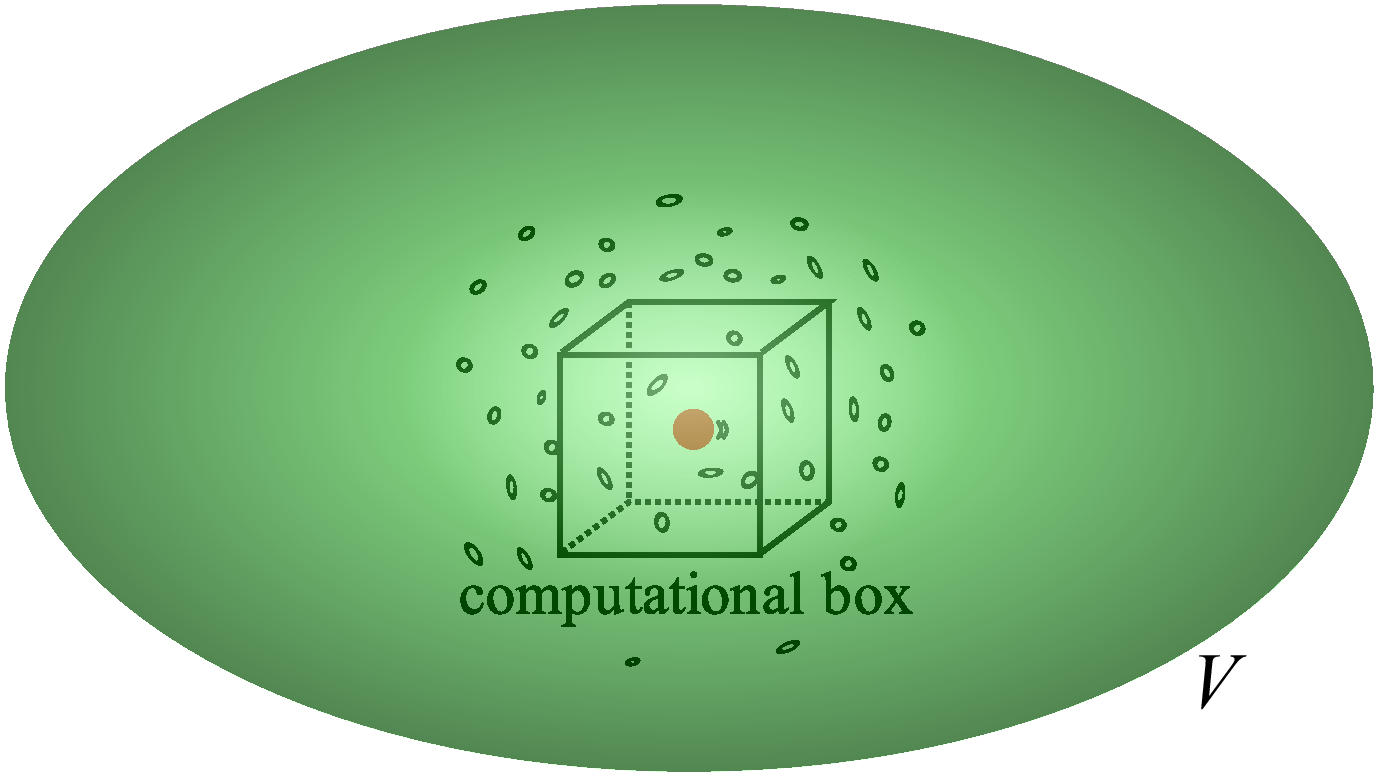}
	\caption{(Color online) Schematic view of the computational box and volume $V$.}\label{fig:box}
\end{figure}

By considering the energy given to the fluid, the drag force can be estimated.
The derivative of the kinetic energy $K$ of the fluid with respect to time is given by
\begin{equation}
	\frac{dK}{dt} = \frac{d}{dt} \int_V \frac{1}{2} \rho_s {v_s}^2 \, dV, \label{eqn:dKdt1}
\end{equation}
where the volume $V$ surrounds the sphere and is chosen to be large enough so that the superfluid velocity vanishes on the outer surface of the volume (Fig.\ \ref{fig:box}).\cite{box}
Note that the derivative and the integral are not interchangeable because the domain of integration changes every time the sphere moves.
To handle the difficulty, we extend the domain of integration to inside the sphere and make the domain time independent.
Instead we define the density of the fluid as a function of time and position,
\begin{equation}
	\rho (\bm{r},t) = \rho_s \theta [|\bm{r}-\bm{x}(t)|-a], \label{eqn:density}
\end{equation}
where $\theta(x)$ is a step function, $\bm{x}(t)$ is the central position of the sphere, and $a$ is the radius of the sphere.
From Eq.\ (\ref{eqn:density}) and Euler's equation, Eq.\ (\ref{eqn:dKdt1}) becomes
\begin{eqnarray}
	\frac{dK}{dt} &=& \int_{V_\mathrm{all}} \frac{1}{2}{v_s}^2 \frac{\partial \rho}{\partial t} \, dV
		+\int_{V_\mathrm{all}} \rho(\bm{r},t) \bm{v_s} \cdot \frac{\partial \bm{v_s}}{\partial t} \, dV \nonumber \\
	& & \nonumber \\
	&=& \int_{V_\mathrm{all}} \frac{1}{2}{v_s}^2 \frac{\partial \rho}{\partial t} \, dV+\int_V \bm{v_s} \cdot \rho_s \frac{\partial \bm{v_s}}{\partial t} \, dV \nonumber \\
	& & \nonumber \\
	&=& \int_{V_\mathrm{all}} \frac{1}{2}{v_s}^2 \frac{\partial \rho}{\partial t} \, dV \nonumber \\
	& & \nonumber \\
	& &+\int_V \bm{v_s}\left\{-\bm{\nabla} p -\rho_s(\bm{v_s}\cdot \bm{\nabla} )\bm{v_s}\right\} \, dV \nonumber \\
	& & \nonumber \\
	&=& \int_{V_\mathrm{all}} \frac{1}{2}{v_s}^2 \frac{\partial \rho}{\partial t} \, dV-\int_V \rho_s \bm{v_s}(\bm{v_s}\cdot \bm{\nabla} )\bm{v_s} \, dV \nonumber \\
	& & \nonumber \\
	& & -\int_V \bm{v_s} \cdot \bm{\nabla} p \, dV, \label{eqn:dKdt2}
\end{eqnarray}
where the volume $V_\mathrm{all}$ is the sum of $V$ and the volume inside the sphere.

We derive that the first and second terms on the right-hand side of Eq.\ (\ref{eqn:dKdt2}) are canceled out.
The time derivative of the integrand in the first term becomes
\begin{eqnarray}
	\frac{\partial \rho}{\partial t} &=& \rho_s \frac{\partial}{\partial t} \theta [|\bm{r}-\bm{x}(t)|-a] \nonumber \\
	& & \nonumber \\
	&=& \rho_s \delta[|\bm{r}-\bm{x}(t)|-a] (\bm{\nabla_x}| \bm{r}-\bm{x}(t)|) \cdot \frac{d\bm{x}(t)}{dt} \nonumber \\
	& & \nonumber \\
	&=& -\rho_s \delta[|\bm{r}-\bm{x}(t)|-a] \frac{\bm{r}-\bm{x}(t)}{|\bm{r}-\bm{x}(t)|} \cdot \bm{u_p}(t), \label{eqn:time-derivative}
\end{eqnarray}
where $\bm{\nabla_x}$ is the gradient with respect to $\bm{x}$.
The delta function in Eq.\ (\ref{eqn:time-derivative}) transforms the integral of the first term on the right-hand side of Eq.\ (\ref{eqn:dKdt2}) into the surface integral,
\begin{eqnarray}
	\int_{V_\mathrm{all}} \frac{1}{2}{v_s}^2 \frac{\partial \rho}{\partial t} \, dV &=& -\int_{V_\mathrm{all}} \frac{1}{2} \rho_s {v_s}^2\delta (|\bm{r}-\bm{x}|-a) \nonumber \\
	& & \nonumber \\
	& & \times \frac{\bm{r}-\bm{x}}{|\bm{r}-\bm{x}|} \cdot \bm{u_p} \, dV \nonumber \\
	& & \nonumber \\
	&=& -\int_S \frac{1}{2}\rho_s {v_s}^2 \bm{u_p}\cdot d\bm{S_1}, \nonumber
\end{eqnarray}
where the area $S$ is the surface of the sphere, and $d\bm{S_1}$ points outside the sphere.

On the other hand, the incompressibility $\bm{\nabla} \cdot \bm{v_s} = 0$ leads to $\bm{v_s}(\bm{v_s} \cdot \bm{\nabla}) \bm{v_s} = \bm{\nabla} (\frac{1}{2} {v_s}^2 \bm{v_s})$ and allows us to rewrite the second term on the right-hand side of Eq.\ (\ref{eqn:dKdt2})
\begin{eqnarray}
	\int_V \rho_s \bm{v_s}(\bm{v_s}\cdot \bm{\nabla} )\bm{v_s} \, dV
	&=& \int_{V} \bm{\nabla} \cdot \left( \frac{1}{2} \rho_s {v_s}^2 \bm{v}\right) dV \nonumber \\
	& & \nonumber \\
	&=& \int_{S'} \rho_s \frac{1}{2}{v_s}^2 \bm{v_s} \cdot d\bm{S}, \nonumber
\end{eqnarray}
where $S'$ is the sum of $S$ and the outer surface of $V$.
Since the velocity on the outer surface of $V$ vanishes, only the integral over the sphere remains, and one obtains
\begin{equation}
	\int_V \rho_s \bm{v_s}(\bm{v_s}\cdot \bm{\nabla} )\bm{v_s} \, dV
	= -\int_S \frac{1}{2} \rho_s {v_s}^2 \bm{u_p}\cdot d\bm{S_1}.
\end{equation}
Thus the first and second terms on the right-hand side of Eq.\ (\ref{eqn:dKdt2}) are canceled out.
Then Eq.\ (\ref{eqn:dKdt2}) becomes
\begin{eqnarray}
	\frac{dK}{dt} &=& -\int_V \bm{v_s} \cdot \bm{\nabla} p \, dV \nonumber \\
	 & & \nonumber \\
	 &=& -\int_V \bm{\nabla} \cdot (p\bm{v_s}) \, dV \nonumber \\
	 & & \nonumber \\
	 &=& -\int_S p\bm{v_s}\cdot d\bm{S_2}, \label{eqn:dKdt3}
\end{eqnarray}
where the incompressibility is used, and $d\bm{S_2}$ points inside the sphere.
From Eq.\ (\ref{eqn:boundary_with_sphere}), $\bm{v_s} \cdot d\bm{S_2} = \bm{u_p} \cdot d\bm{S_2}$, and Eq.\ (\ref{eqn:dKdt3}) becomes
\begin{equation}
	\frac{dK}{dt} = -\int _S p \bm{u_p} \cdot d\bm{S_2} = -\bm{u_p} \cdot \left(\int _S p \, d\bm{S_2} \right). \label{eqn:dKdt4}
\end{equation}
In Eq.\ (\ref{eqn:dKdt4}), the integral equals the force $\bm{F}_{f \rightarrow s} $ exerted by the fluid on the sphere, and one obtains
\begin{equation}
	\frac{dK}{dt} = -\bm{u_p} \cdot \bm{F}_{f \rightarrow s}. \label{eqn:dKdt5}
\end{equation}
The force $\bm{F}_{f \rightarrow s}$ is resolved into the drag force $\bm{F}_\mathrm{drag}$ parallel to $\bm{u_p}$ and the lift force $\bm{F}_\mathrm{lift}$ perpendicular to $\bm{u_p}$, and finally Eq.\ (\ref{eqn:dKdt5}) becomes
\begin{equation}
	\frac{dK}{dt} = -\bm{u_p} \cdot \bm{F}_\mathrm{drag}. \label{eqn:dKdt6}
\end{equation}

Here we assume that $\bm{F}_\mathrm{drag}$ is antiparallel to $\bm{u_p}$ averaged over a time period $T_\mathrm{int}$ much longer than the oscillation period, and that $\bm{u_p}$ and $\bm{F}_\mathrm{drag}$ oscillate in phase.
The first assumption is based on the fact that a drag force giving energy to the surrounding fluid works against motion of an object.
For the second assumption, it is considered that the drag force works when the vortices on the sphere are stretched, and the numerical simulation of Fig.\ \ref{fig:turbulence} clearly shows that the vortex stretch is caused by the motion of the sphere.\cite{movie}
As a result, the drag force is synchronized with the oscillation of the sphere through the vortex growth without delay, and the second assumption is found to be valid.
Then, taking the average of Eq.\ (\ref{eqn:dKdt6}) over $T_\mathrm{int}$, we obtain
\begin{equation}
	\left\langle \frac{dK}{dt} \right\rangle
	= \langle -\bm{u_p} \cdot \bm{F}_\mathrm{drag} \rangle
	\cong \, u_p^0 \cdot F_\mathrm{drag}^0 \, ,
\end{equation}
where $u_p^0$ and $F_\mathrm{drag}^0$ are the amplitude of each oscillating quantity.
Consequently the drag force is evaluated by dividing $\langle dK/dt \rangle$ by the velocity of the sphere.
In the equilibrium state of quantum turbulence, the increase in the kinetic energy of the fluid in volume $V$ equals to the energy of the vortices that escape from the computational box.\cite{box}
Thus $\langle dK/dt \rangle$ can be evaluated by averaging the energy of vortices that escape from the box over time period $T_\mathrm{int}$.
The energy of vortices can approximately be evaluated from its total line length.
The energy of the vortex per unit length is given by
\begin{equation}
	\epsilon = \frac{\rho_s \kappa^2}{4\pi} \ln \left( \frac{L}{a_0} \right), \label{eqn:unit_energy}
\end{equation}
where $L$ is the characteristic length in the system, namely the computational box in this case.
The line length of the vortices escaping from the box is measured over $T_\mathrm{int}$ = 100 ms, and the energy is calculated by multiplying Eq.\ (\ref{eqn:unit_energy}) by the length.\cite{interval}
The line length also includes that of the injected vortices so the length of the injected vortices is subtracted from that of vortices that escape from the box.
\begin{figure}[tb]
	\begin{minipage}{\linewidth}
		\includegraphics[width=0.49\linewidth]{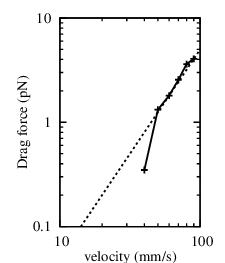}
		\includegraphics[width=0.49\linewidth]{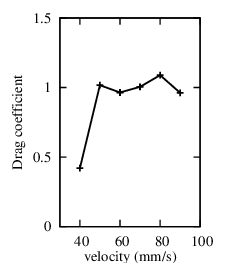}
	\end{minipage}
	\caption{The drag force and drag coefficient as a function of the velocity.
		The dotted line on the left-hand figure indicates the slope of $v^2$.}\label{fig:drag_force}
\end{figure}

The drag force is plotted in Fig.\ \ref{fig:drag_force} as a function of the oscillation velocity.
It should be noted that the unit of the drag force is pN, which is hundredths of the previously presented experimental data, say, 0.1 nN in Ref.\ 10 and 0.05 nN in Ref.\ 14.
This is explained by the difference in the object sizes used in the present study in comparison with these previous experimental studies.
In Ref.\ 10, J\"{a}ger \textit{et al.}\ used an oscillating sphere with radius 100 $\mu$m, which is a hundred times larger than ours, and accordingly the vortex tangle created around the sphere would be larger than that of the present results.
In Ref.\ 14, the length of wire over which vortex tangle is created is not known because the legs of the wire are installed on the experimental cell and only the middle of it can vibrate.
Let us suppose a tenth of the wire effectively contributes to the turbulence.
The diameter of the wire is 2.5 $\mu$m and the whole length is 2 mm (then 200 $\mu$m of the wire creates the turbulence), and the ratio of the wire size to the sphere is about 60.
It is expected that the wire could create at least 50 times of the size of the vortex tangle of our results.
Hence, the observed difference in magnitudes of the drag force found by the present study and the previous experimental observations may be reasonable.

In the range of velocity magnitudes from 50 to 90 mm/s, the drag force agrees well with a $v^2$ profile.
Substituting the values of the drag force to Eq.\ (\ref{eqn:drag_force}) yields the drag coefficient, and such results are plotted in Fig.\ \ref{fig:drag_force}.
The order of the drag coefficient in Eq.\ (\ref{eqn:drag_force}) is unity, being consistent with the values found  in the related experiments.\cite{Vinen}

Skrbek and Vinen discussed in Ref.\ 17 that quantum turbulence created by oscillating objects at zero temperature also exhibits features similar to those of classical counterpart by citing various experiments.
As we mentioned in Sec.\ \ref{introduction}, the similarity between these phenomena is confirmed in homogeneous and isotropic turbulence, as the form of Kolmogorov's law.
It is remarkable that this similarity is numerically confirmed in another type of turbulence, say, turbulence created by oscillating objects.

Skrbek and Vinen\cite{Vinen}  also described the existence of two possibilities for the flow of the superfluid. One is \textit{quasiclassical laminar flow}, in which a vortex tangle of low density is formed around the objects and the flow of the superfluid mimics a classical laminar flow. The other is \textit{quasiclassical turbulent flow}, in which the density of vortex tangle becomes high and large scale rotational flow appears, mimicking classical turbulence.
From Fig.\ \ref{fig:turbulence}, vortices of large scale comparable to the sphere size can be seen, and in fact, the flow exhibits the behavior of Eq.\ (\ref{eqn:drag_force}).
These agreements evince the similarity between the classical and quantum turbulences and support the picture of quasiclassical turbulent flow.

The process whereby the turbulence is created raises the possibility that the time interval $\tau$ in which the vortex rings are injected can affect the drag coefficient.
\begin{table}[t]
	\caption{The drag coefficient computed for different time intervals. Oscillation velocity is 50 mm/s.}\label{tab:Cd}
\begin{ruledtabular}
\begin{tabular}{llll}
	$\tau$(ms)	&	0.03	&	0.05	&	0.1 \\
	$C_D$			&	0.78	&	0.68	&	0.59 \\
\end{tabular}
\end{ruledtabular}
\end{table}
Table \ref{tab:Cd} shows the drag coefficient calculated using different time intervals at an oscillation velocity of magnitude 50 mm/s.
As the injection interval becomes shorter, the rate of vortex growth increases, resulting in an increase in the drag coefficient.
However, it may be noted that the drag coefficient remains of order unity.

\section{conclusion and discussion}\label{conclusion}
Quantum turbulence was created around an oscillating sphere, and its evolution was studied.
We proposed a method to evaluate the drag force in quantum turbulence.
The drag force was calculated by considering the energy supplied by the sphere, and it is proportional to the square of the magnitude of the oscillation velocity.
This dependency on the velocity is quite similar to that in classical turbulence.
The drag coefficient was also calculated, and its order of magnitude is in agreement with that of the previous experimental results.
The similarity between classical and quantum turbulences was confirmed by numerical simulations.

We believe that the drag coefficient has an universal value independent of the details of the system.
Our future work is to confirm whether the drag coefficient still remains of the order of unity even if the parameters such as the frequency of the oscillation and the geometry of the oscillating object are changed.

The simulations were performed for a smooth sphere, but the objects used in the experiments have surface roughness.
The roughness may increase the drag force and the drag coefficient;
when a vortex is detached from the sphere due to the oscillation, it is likely to leave a small vortex bridge over the pinning sites on the surface.
The small vortex plays a role of ``seed'' for the growth of the vortex, and this is equivalent to injecting more vortices to the sphere, resulting in an increase in the drag force and the drag coefficient.
However, there is no clear way to proceed to obtain the solution of the equations of motion of vortices under the boundary conditions which describe the surface roughness at present, and the numerical simulations require a model that phenomenologically depicts the vortex pinning.

Figure \ref{fig:drag_force} enables us to discuss the transition from the turbulent to laminar state.
Below oscillation velocities of magnitude 50 mm/s, the drag force starts to deviate from Eq.\ (\ref{eqn:drag_force}) and tends to zero.
This is because the motion of the sphere is so slow that vortices cannot be stretched before the attached vortices leave the sphere, resulting in the drag force tending to zero.
This means that the oscillating sphere can no longer sustain turbulence, and that the flow regime will return to a laminar state.
From dimensional analysis, the critical velocity at which turbulence returns to laminar flow is given by $v_c = c \sqrt{\kappa \omega}$, where $c$ is constant with the order of unity, and this formulation agrees with the critical velocity in the experiments.\cite{Hanninen2}
In the simulation, $\kappa = 10^{-7}$ m$^2$/s, $\omega = 10^4$ rad/s, and finally one obtains $v_c \sim$ 30 mm/s.
This is very close to the velocity 50 mm/s below which the deviation from Eq.\ (\ref{eqn:drag_force}) starts in the simulation.
Therefore the velocity magnitude 50 mm/s can be considered in the simulations to represent  the critical velocity magnitude.

\begin{acknowledgments}
S.F.\ is very grateful for helpful discussion with Risto H\"{a}nninen.
M.T.\ acknowledges the supports of Grant-in-Aid for Scientific Research from JSPS (Grant No.\ 18340109) and Grant-in-Aid for Scientific Research on Priority Areas (Grant No.\ 17071008) from MEXT.
\end{acknowledgments}


\begin{thebibliography}{30}
	\bibitem{Tsubota}M. Tsubota,
		J. Phys. Soc. Jpn. \textbf{77}, 111006 (2008).
	\bibitem{PLTP}
		\textit{Progress in Low Temperature Physics}, edited by W. P. Halperin and M. Tsubota (North-Holland, Amsterdam, 2008), Vol.\ 16.
	\bibitem{Donnelly}R. J. Donnelly,
		\textit{Quantized Vortices in Helium II} (Cambridge University Press, Cambridge, 1991).
	\bibitem{Hashimoto}N.\ Hashimoto, R. Goto, H. Yano, K. Obara, O. Ishikawa, and T. Hata,
		Phys. Rev. B \textbf{76}, 020504(R) (2007).
	\bibitem{Nore}
		C. Nore, M. Abid, and M. E. Brachet,
		Phys. Rev. Lett. \textbf{78}, 3896 (1997).
	\bibitem{Maurer}
		J. Maurer and P. Tabeling,
		Europhys. Lett. \textbf{43}, 29 (1998).
    \bibitem{Stalp}
		S. R. Stalp, L. Skrbek, and R. J. Donnelly,
		Phys. Rev. Lett. \textbf{82}, 4831 (1999).
	\bibitem{Araki1}
		T. Araki, M. Tsubota, and S. K. Nemirovskii,
		Phys. Rev. Lett. \textbf{89}, 145301 (2002).
	\bibitem{Kobayashi}
		M. Kobayashi and M. Tsubota,
		Phys. Rev. Lett. \textbf{94}, 065302 (2005)
		J. Phys. Soc. Jpn. \textbf{74}, 3248 (2005).
	\bibitem{Schoepe}J. J\"{a}ger, B. Schuderer, and W. Schoepe,
		Phys. Rev. Lett. \textbf{74}, 566 (1995).
	\bibitem{Fisher}S. N. Fisher, A. J. Hale, A. M. Gu\'{e}nault, and G. R. Pickett,
		Phys. Rev. Lett. \textbf{86}, 244 (2001).
	\bibitem{McClintock}H. A. Nichol, L. Skrbek, P. C. Hendry, and P. V. E. McClintock,
		Phys. Rev. Lett. \textbf{92}, 244501 (2004).
	\bibitem{Bradley}D. I. Bradley, D. O. Clubb, S. N. Fisher, A. M. Gu\'enault, R. P. Haley, C. J. Matthews, G. R. Pickett, V. Tsepelin, and K. Zaki,
		Phys. Rev. Lett. \textbf{95}, 035302 (2005).
	\bibitem{Yano}H. Yano, N. Hashimoto, A. Handa, M. Nakagawa, K. Obara, O. Ishikawa, and T. Hata,
		Phys. Rev. B \textbf{75}, 012502 (2007).
	\bibitem{Goto}R. Goto, S. Fujiyama, H. Yano, Y. Nago, N. Hashimoto, K. Obara, O. Ishikawa, M. Tsubota, and T. Hata,
		Phys. Rev. Lett. \textbf{100}, 045301 (2008).
	\bibitem{Blazkova}M. Bla\v{z}kov\'{a}, M. \v{C}love\v{c}ko, V. B. Eltsov, E. Ga\v{z}o, R. de Graaf, J. J. Hosio, M. Krusius, D. Schmoranzer, W. Schoepe, L. Skrbek, P. Skyba, R. E. Solntsev, and W. F. Vinen,
		J. Low Temp. Phys. \textbf{150}, 525 (2008).
	\bibitem{Vinen}L. Skrbek and W. F. Vinen, in \textit{Progress in Low Temperature Physics}, edited by W. P. Halperin and M. Tsubota (North-Holland, Amsterdam, 2008), Vol. 16, p. 193.
	\bibitem{Kivotides}D. Kivotides, C. F. Barenghi, and Y. A. Sergeev,
		Phys. Rev. B \textbf{75}, 212502 (2007) \textbf{77}, 014527 (2008).
	\bibitem{Hanninen1}R. H\"anninen, M. Tsubota, and W. F. Vinen,
		Phys. Rev. B \textbf{75}, 064502 (2007).
	\bibitem{Schwarz85}K. W. Schwarz,
		Phys. Rev. B \textbf{31}, 5782 (1985).
	\bibitem{Schwarz88}K. W. Schwarz,
		Phys. Rev. B \textbf{38}, 2398 (1988).
	\bibitem{Saffman}P. G. Saffman,
		\textit{Vortex Dynamics} (Cambridge University Press, Cambridge, 1992).
	\bibitem{Laplace_solve}
		An alternative solution can also be used and is valid even for a vortex ending at boundaries that extend infinitely. See Ref. 24. However, the method of the image vortex reduces the required computational time and is advantageous for this calculation.
	\bibitem{Schwarz74}K. W. Schwarz,
		Phys. Rev. A \textbf{10}, 2306 (1974).
	\bibitem{Koplik}J. Koplik and H. Levine,
		Phys. Rev. Lett. \textbf{71}, 1375 (1993).
	\bibitem{Araki2}M. Tsubota, T. Araki, and S. K. Nemirovskii,
		Phys. Rev. B \textbf{62}, 11751 (2000).
	\bibitem{movie}See EPAPS Document No. E-PRMBDO-79-010909 for the movie of turbulence generation for the case of a sphere oscillating with a velocity magnitude of  90 mm/s. For more information on EPAPS, see http://www.aip.org/pubservs/epaps.html
	\bibitem{Landau}L. D. Landau and E. M. Lifschitz,
		\textit{Fluid Mechanics},2nd ed. (Pergamon, London, 1987).
	\bibitem{box}Note that the computational box is different from $V$.
		The volume $V$ is considered large enough so that the vortices never escape from $V$ in the time period $T_\mathrm{int}$, while vortices can escape from the computational box within $T_\mathrm{int}$.
	\bibitem{interval}In the time interval $T_\mathrm{int}$= 100 ms, many vortices escape from the computational box as easily found in Fig.\ \ref{fig:length}, and the statistical fluctuation of $\langle dK/dt \rangle$ is supposed to be small, we have not yet estimated the fluctuation of $\langle dK/dt \rangle$ systematically though.
	\bibitem{Hanninen2}R. H\"anninen and W. Schoepe,
		J. Low Temp. Phys. \textbf{153}, 189 (2008).
\end{thebibliography}
\end{document}